\documentclass[10pt, conference]{IEEEtran}
\usepackage{amsmath}
\usepackage{amssymb}
\usepackage{amsfonts}
\usepackage{amscd}
\usepackage{mathrsfs}
\usepackage[final]{graphicx}
\usepackage{graphicx}
\usepackage{color}
\usepackage{url}

\newtheorem{theorem}{Theorem}
\newtheorem{lemma}{Lemma}

\newtheorem{proposition}{Proposition}

\begin{document}

\title{Trellis Coded Modulation for Two-User Unequal-Rate Gaussian MAC}

\author{
\authorblockN{J. Harshan}
\authorblockA{Dept. of ECE, Indian Institute of science \\
Bangalore 560012, India\\
Email:harshan@ece.iisc.ernet.in\\
}
\and
\authorblockN{B. Sundar Rajan}
\authorblockA{Dept. of ECE, Indian Institute of science \\
Bangalore 560012, India\\
Email:bsrajan@ece.iisc.ernet.in\\
}
}
\maketitle

\begin{abstract}
In this paper, code pairs based on trellis coded modulation are proposed over PSK signal sets for a two-user Gaussian multiple access channel. In order to provide unique decodability property to the receiver and to maximally enlarge the constellation constrained (CC) capacity region, a relative angle of rotation is introduced between the signal sets. Subsequently, the structure of the \textit{sum alphabet} of two PSK signal sets is exploited to prove that Ungerboeck labelling on the trellis of each user maximizes the guaranteed minimum squared Euclidean distance, $d^{2}_{g, min}$ in the \textit{sum trellis}. Hence, such a labelling scheme can be used systematically to construct trellis code pairs for a two-user GMAC to approach \emph{any rate pair} within the capacity region.
\end{abstract}


\section{Introduction and Preliminaries}
\label{sec1}
\begin{figure}
\centering
\includegraphics[width=2.3in]{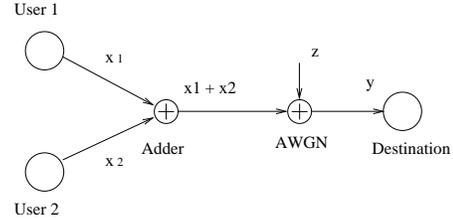}
\caption{Two-user Gaussian MAC model} 
\label{gmac_model}
\end{figure}

\begin{figure*}
\centering
\includegraphics[width=6in]{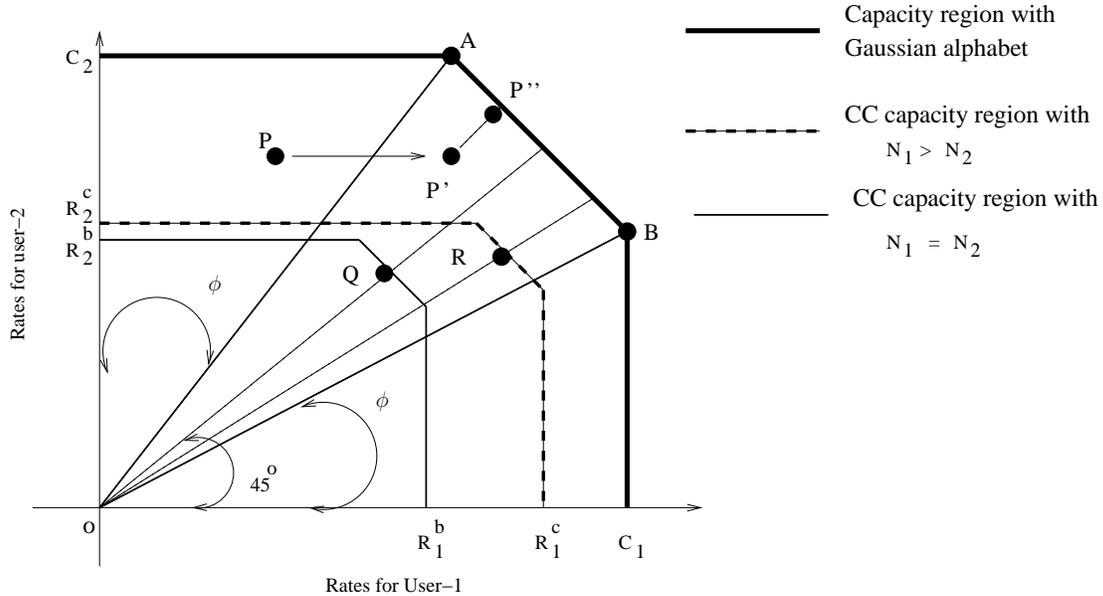}
\caption{Capacity regions of two-user GMAC (for a fixed equal average power constraint) with (i) finite alphabets and (ii) Gaussian alphabets.} 
\label{CC_region_equal}
\end{figure*}

Works on coding for Gaussian multiple access channel with finite complex alphabets have been reported in \cite{FTL}-\cite{WCA}. Constellation Constrained (CC) capacity regions \cite{Eb} of a two-user GMAC (shown in Fig. \ref{gmac_model}) have been computed in \cite{HaR1} wherein the unique decodability (UD) property at the destination is achieved by appropriate rotation between the alphabets of the two users. The angles of rotation between the alphabets which enlarge the CC capacity regions have also been computed.\\ 
\indent For a fixed equal average power constraint for the two users, the Gaussian capacity region of two-user GMAC is shown in Fig. \ref{CC_region_equal} along with the CC capacity regions of two pairs of finite alphabets. In Fig. \ref{CC_region_equal}, the average power constraints for all the three cases are equal and, hence, the Gaussian capacity region encloses the other two regions. For the channel model in \cite{HaR1}, code pairs based on trellis coded modulation (TCM) \cite{Ub} are proposed in \cite{HaR2} such that sum-rates close to the sum capacity can be achieved. In such a setup, the number of information bits transmitted by the two users are equal. In other words, coding schemes in \cite{HaR2} can potentially approach any rate pair only on the $45$ degrees line within the capacity region shown in Fig. \ref{CC_region_equal} (for example, the point Q). Note that, using the set up in \cite{HaR2}, rate pairs other than those on the $45$ degrees line in Fig. \ref{CC_region_equal} cannot be achieved (for example, the point R or the point P shown in Fig. \ref{CC_region_equal} cannot be achieved).\\
\indent In this paper, we propose TCM based trellis codes to approach \emph{any rate pair} (for example, the points R, P and P' shown in Fig. \ref{CC_region_equal}) within the capacity region of a two-user GMAC with equal power constraint for the two users. In other words, unlike \cite{HaR2}, the setup here permits the two users to transmit unequal number of information bits to the destination. The proposed codes are applicable when the two users have PSK signal sets. Since the signal sets can have unequal size and equal average power constraint, the corresponding CC capacity region can be asymmetric. For example, see Fig. \ref{CC_region_equal} which shows the CC capacity region of unequal sized signal sets with $N_{1} > N_{2}$ (where $N_{1}$ and $N_{2}$ are the sizes of signal sets of User-1 and User-2 respectively).\\
\indent Throughout the paper, the terms alphabet and signal set are used interchangeably. All the rates mentioned in this paper are in bits per channel use. Unless specified, whenever $N_{1}$ and $N_{2}$ are unequal, we assume $N_{2} \geq  N_{1}$ throughout the paper. The contributions of the paper may be summarized as below: 
\begin{itemize}
\item For a two-user GMAC with PSK signal sets and equal average power constraints for the two users, we use a generalised version of the metric in \cite{HaR1}, \cite{HaR3} to compute relative angle(s) of rotation $\theta$, between the signal sets to provide maximum enlargement of the CC capacity region. In particular, when User-$1$ and User-$2$ are equipped with $N_{1}$-PSK and $N_{2}$-PSK respectively, it is shown that, $\theta = \frac{\pi}{N_{2}}$ between the alphabets maximally enlarges the CC capacity region (Section \ref{sec3}).\\
\item For each $i = 1 \mbox{ and } 2$, if User-$i$ employs the trellis $T_{i}$ labelled with the symbols of $N_{i}$-PSK, it is clear that the destination views the sum trellis, $T_{sum}$  labelled with the symbols of the sum alphabet, $\mathcal{S}_{sum}$ of $N_{1}$-PSK and $N_{2}$-PSK signal sets in an equivalent SISO (Single-Input Single-Output) AWGN channel \cite{HaR2}. When the angle, $\theta$ is  $\frac{\pi}{N_{2}}$, it is shown that, Ungerboeck labelling on the trellis of each user induces a labelling on $T_{sum}$ which maximizes the guaranteed minimum squared Euclidean distance, $d^{2}_{g, min}$ of $T_{sum}$ (See Theorem \ref{ungerboeck_theorem} in Section \ref{sec3_subsec3}). Hence, for a given power constraint and a given pair ($N_{1}, N_{2}$), the proposed labelling scheme can be used systematically to construct trellis code pairs to approach any rate pair on the line inclined at $\mbox{tan}^{-1}(\frac{\mbox{log}_{2}(N_{2})}{\mbox{log}_{2}(N_{1})})$ degrees with the x-axis, within CC capacity region. As an example, for the CC capacity region shown in Fig. \ref{CC_region_equal}, any rate pair on the line joining the origin and the point R can be achieved. \\
\item We focus on approaching rate pairs on the line segment A-B shown in Fig. \ref{CC_region_equal} since such rate pairs maximize the sum capacity. For a given rate pair, say point P' within the sector O-A-B shown in Fig. \ref{CC_region_equal}, we propose a method to choose a sequence of ($N_{1}$, $N_{2}$) pairs which when employed in our coding scheme can approach any rate pair on the line segment P'-P'' (Section \ref{sec3_subsec4}).
\end{itemize}
\indent The above contributions along with \cite{HaR1} \cite{HaR2} and \cite{HaR3} establish that, whenever the input alphabets are PSK signal sets and the average power constraints for both the users are equal, explicit practical TCM schemes approaching any rate pair within the capacity region can be obtained for two-user Gaussian multiple access channels.

\textit{Notations:} Cardinality of the set $\mathcal{S}$ is denoted by $|\mathcal{S}|$. Absolute value of a complex number $x$ is denoted by $|x|$ and $E \left[x\right]$ denotes the expectation of the random variable $x$. A circularly symmetric complex Gaussian random vector, $\textbf{x}$ with mean $\mathbf{\mu}$ and covariance matrix $\Gamma$ is denoted by $\textbf{x} \sim \mathcal{CN} \left(\mu, \Gamma \right)$. Also, the set of all real and complex numbers are denoted by $\mathbb{R}$ and $\mathbb{C}$ respectively. For two points $a,b \in \mathbb{C}$, $d(a, b)$ denotes the Euclidean distance between $a$ and $b$.

\indent The remaining content of the paper is organized as follows: In Section \ref{sec2}, the two-user GMAC model along with the description of the coding problem is presented. In Section \ref{sec3}, details on designing TCM based code pairs with a pair of unequal sized PSK signal sets are presented. Section \ref{sec4} constitutes conclusion and some directions for possible further work.
\section{Trellis Coded Modulation (TCM) for a two-user GMAC}
\label{sec2}
\begin{figure}
\centering
\includegraphics[width=3.3in]{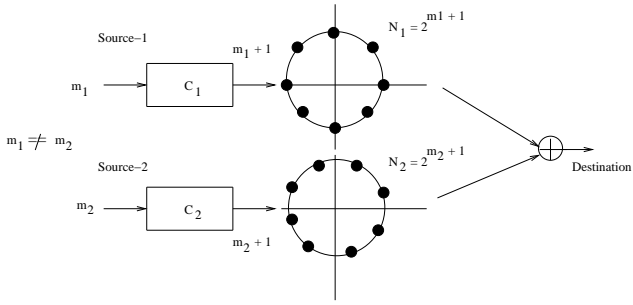}
\caption{TCM scheme which adds equal number of redundancy bits for both the users to handle unequal rates} 
\label{scheme_1}
\end{figure}

The model of a two-user Gaussian MAC as shown in Fig. \ref{gmac_model} consists of two users that need to convey information to a single destination. It is assumed that User-1 and User-2 communicate to the destination at the same time and in the same frequency band. Symbol level synchronization is assumed at the destination. The two users are equipped with alphabets $\mathcal{S}_{1}$ and $\mathcal{S}_{2}$ of size $N_{1}$ and $N_{2}$ respectively. When User-1 and User-2 transmit symbols $x_{1} \in \mathcal{S}_{1}$ and $x_{2} \in \mathcal{S}_{2}$ respectively, the destination receives a symbol $y$ given by,
\begin{equation}
y = \rho x_{1} + \rho x_{2} + z ~~ \mbox{where }  z \sim \mathcal{CN} \left(0, \sigma^{2} \right).
\end{equation}
where $\rho$ is the power used by each user for every channel use. Henceforth, unless specified, we assume $\rho = 1$ for simplicity. For the above channel, in \cite{HaR1}, CC capacity regions have been computed for some well known alphabets such as $M$-PSK, $M$-QAM etc and the angles of rotation between the alphabets that maximally enlarges the CC capacity region have also been presented. In this paper, we design code pairs $(\mathcal{C}_{1}, \mathcal{C}_{2})$ over PSK signal sets to approach rate pairs, ($m_{1}, m_{2}$) ($m_{1}$ for User-1 and $m_{2}$ for user-2 such that $m_{1}$ and $m_{2}$ can be unequal) within the Gaussian capacity region.\\
\indent For each $i = 1, 2$, let User-$i$ be equipped with a convolutional encoder $C_{i}$ with $m_{i}$ input bits and $m_{i} + 1$ output bits as shown in Fig. \ref{scheme_1}. Throughout the paper, we consider convolutional encoders which add 1 bit of redundancy only. Let the $m_{i} + 1$ output bits of $C_{i}$ take values from a $2$-dimensional signal set $\mathcal{S}_{i}$ such that $N_{i} = 2^{m_{i} + 1}$. Henceforth, the codes $\mathcal{C}_{1}$ (set of codewords generated from $C_{1}$) and $\mathcal{C}_{2}$ (set of codewords generated from $C_{2}$) are represented by trellises $T_{1}$ and $T_{2}$ respectively. The definition of sum trellis, $T_{sum}$ for the trellis pair $\left(T_{1}, T_{2}\right)$ and sum alphabet, $\mathcal{S}_{sum}$ of $\mathcal{S}_{1}$ and $\mathcal{S}_{2}$ have been introduced in \cite{HaR2}.\\
\indent We assume that the destination performs joint decoding of the symbols of User-1 and User-2 by decoding for a sequence over $\mathcal{S}_{sum}$ on $T_{sum}$. For the trellis pair $(T_{1}, T_{2})$ and the alphabet pair $(\mathcal{S}_{1}, \mathcal{S}_{2})$, the destination views an equivalent SISO AWGN channel with a virtual source equipped with the trellis, $T_{sum}$ labelled with the elements of $\mathcal{S}_{sum}$. Due to the existence of an equivalent AWGN channel in the two-user GMAC set-up, $T_{sum}$ has to be labelled with the elements of $\mathcal{S}_{sum}$ satisfying the Ungerboeck design rules \cite{Ub}. However, such a labelling rule can be obtained on $T_{sum}$ only through the pairs $(T_{1}, T_{2})$ and $(\mathcal{S}_{1}, \mathcal{S}_{2})$. Hence, in this paper, we propose labelling rules on $T_{1}$ and $T_{2}$ using $\mathcal{S}_{1}$ and $\mathcal{S}_{2}$ respectively such that $T_{sum}$ is labelled with the elements of $\mathcal{S}_{sum}$ as per Ungerboeck rules.

\indent For each $i = 1 \mbox{ and } 2$, since the number of input bits to $C_{i}$ is $m_{i}$, there are $2^{m_{i}}$ edges originating from (or converging to; henceforth, we only refer to originating edges) each state of $T_{i}$. Also, as there is only one bit redundancy to be added by the encoder and $N_{i} = 2^{m_{i} + 1}$, the edges originating from each state have to be labelled with the elements of a proper subset of $\mathcal{S}_{i}$ of size $2^{m_{i}}$. Therefore, for each $i = 1 \mbox{ and } 2$, $\mathcal{S}_{i}$ has to be partitioned into two sets namely, $\mathcal{S}^{1}_{i}$ and $\mathcal{S}^{2}_{i}$ and the diverging edges from each state of $T_{i}$ have to be labelled with the elements of either $\mathcal{S}^{1}_{i}$ or $\mathcal{S}^{2}_{i}$. From the definition of a sum trellis, there are $2^{m_{1} + m_{2}}$ edges originating from each state of $T_{sum}$ which gets labelled with the elements of one of the following sets,
\begin{equation*}
\mathcal{A} = \left\lbrace \mathcal{S}^{i}_{1} + \mathcal{S}^{j}_{2} ~ \forall ~ i, j = 1, 2 \right\rbrace. 
\end{equation*}
As per the Ungerboeck design rules, the transitions originating from the same state of $T_{sum}$ must be assigned symbols that are separated by largest minimum Euclidean distance. Hence, an optimal partitioning of $\mathcal{S}_{i}$ into two sets $\mathcal{S}^{1}_{i}$ and $\mathcal{S}^{2}_{i}$ of equal cardinality has to be proposed such that the minimum Euclidean distance, $d_{min}$ of each one of the sets in $\mathcal{A}$ is maximized. However, since $d_{min}$ of the sets in $\mathcal{A}$ can potentially be different, we try to find an optimal partitioning such that the minimum of the $d_{min}$ values of the sets in $\mathcal{A}$ is maximized. The minimum of the $d_{min}$ values of the sets in $\mathcal{A}$ is referred as the guaranteed minimum squared Euclidean distance, $d^{2}_{g, min}$ of $T_{sum}$.
\begin{figure*}
\begin{equation}
\label{thmeq}
M(\theta) = \mbox{arg} \min_{\theta \in (0, 2 \pi)} \sum_{k_{1} = 0}^{N_{1} - 1}\sum_{k_{2} = 0}^{N_{2} - 1}\mbox{log}_{2}\left[ \sum_{i_{1} = 0}^{N_{1} - 1}\sum_{i_{2} = 0}^{N_{2} - 1} \mbox{exp}\left(- |x_{1}(k_{1}) - x_{1}(i_{1}) + e^{i \theta} ( x_{1}(k_{2})  - x_{1}(i_{2})) |^{2}/4\sigma^{2}\right)\right] .
\end{equation}
\end{figure*}
\section{TCM schemes with unequal sized PSK signal sets}
\label{sec3}
\indent  In this section, we propose a solution to the problem of designing labelling rules when $\mathcal{S}_{1}$ and $\mathcal{S}_{2}$ are symmetric PSK signal sets of cardinality $N_{1}$ and $N_{2}$ respectively where $N_{1} = 2^{u_{1}}$ and $N_{2} = 2^{u_{2}}$ for some $u_{1}, u_{2} \geq 1$.  Without loss of generality, we assume that $N_{2} \geq N_{1}$. Let $k$ denote the ratio $\frac{N_{2}}{N_{1}}$ (note that $\frac{N_{2}}{N_{1}} \geq 1$). To obtain the unique decodability property at the receiver \cite{HaR1}, we employ a $\theta$-rotated version of $\mathcal{S}_{2}$. 

When the two users are equipped with identical signal sets, a metric has been proposed in Theorem 1 of \cite{HaR1}, to compute relative angle(s) of rotation between the alphabets such that the corresponding CC capacity region is maximally enlarged. However, the proof of Theorem 1 in \cite{HaR1} is not applicable when the cardinality of the signal sets are unequal. Hence, proceeding on the similar lines of the results in Theorem 1 of \cite{HaR1}, a new metric shown in \eqref{thmeq} can be obtained to compute relative angle(s) of rotation between two \textit{unequal} sized alphabets. Using the metric in \eqref{thmeq}, the angle, $\theta = \frac{\pi}{N_{2}}$ can be shown to maximally enlarge the CC capacity region of $N_{1}$-PSK  and $N_{2}$-PSK signal sets. Henceforth, we use $\theta = \frac{\pi}{N_{2}}$ throughout the paper.
\subsection{Structure of the sum alphabet of two unequal sized PSK signal sets}
\label{sec3_subsec1}
\begin{figure}
\centering
\includegraphics[width=2.5in]{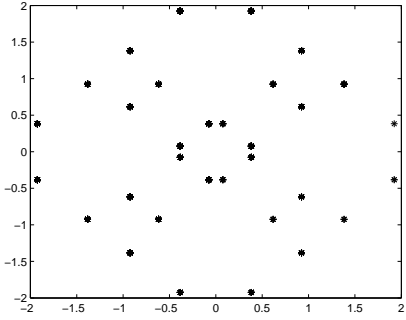}
\caption{The structure of $\mathcal{S}_{sum}$ when $\mathcal{S}_{1}$ = QPSK and $\mathcal{S}_{2}$ = $8$-PSK when $\theta = \frac{\pi}{8}$} 
\label{sum_alphabet_pi_M}
\end{figure}

\indent Let $\mathcal{S}_{1}$ and $\mathcal{S}_{2}$ represent two symmetric PSK signal sets of cardinality $N_{1}$ and $N_{2}$ respectively. The signal set $\mathcal{S}_{2}$ is rotated by an angle $\frac{\pi}{N_{2}}$. Let $x(n_{1})$ and $x'(n_{2})$ denote the points $e^{\frac{i2\pi n_{1}}{N_{1}}}$ and $e^{\frac{i2\pi n_{2}}{N_{2}}}e^{\frac{i \pi}{N_{2}}}$ of $\mathcal{S}_{1}$ and $\mathcal{S}_{2}$ respectively for $ 0 \leq n_{1} \leq N_{1} - 1$ and $ 0 \leq n_{2} \leq N_{2} - 1$. The sum alphabet, $\mathcal{S}_{sum}$ of $\mathcal{S}_{1}$ and $\mathcal{S}_{2}$ can be written as given in \eqref{sum_alph_alt} (at the top of the next page)
\begin{figure*}
\begin{equation}
\label{sum_alph_alt}
\mathcal{S}_{sum} =  \left\lbrace x(n_{1}) + x'(kn_{1} + m), x(n_{1}) + x'(kn_{1} - m - 1) ~|~ \forall ~ 0 \leq n_{1} \leq N_{1} - 1 \mbox{ and } 0 \leq m \leq \frac{N_{2}}{2} - 1 \right\rbrace
\end{equation}
\hrule
\end{figure*}
where 
\begin{equation*}
x(n_{1}) + x'(kn_{1} + m) = e^{\frac{i2\pi n_{1}}{N_{1}}} + e^{i\{\frac{2\pi n_{1}}{N_{1}} + \frac{\pi(2m+1)}{N_{2}}\}}
\end{equation*}
and 
\begin{equation*}
x(n_{1}) + x'(kn_{1} - m - 1) = e^{\frac{i2\pi n_{1}}{N_{1}}} + e^{i\{\frac{2\pi n_{1}}{N_{1}} - \frac{\pi(2m+1)}{N_{2}}\}}
\end{equation*}
such that $x'(-p) = x'(N_{2}-p)$ for any $0 \leq p \leq N_{2}-1$. The phase components of the points $x(n_{1}) + x'(kn_{1} + m)$ and $x(n_{1}) + x'(kn_{1} - m - 1)$ are given by $\frac{2\pi n_{1}}{N_{1}} + \frac{\pi (2m+1)}{2N_{2}}$ and  $\frac{2\pi n_{1}}{N_{1}} - \frac{\pi (2m+1)}{2N_{2}}$ respectively. For a fixed $m$, the set of points of the form $x(n_{1}) + x'(kn_{1} + m)$ and $x(n_{1}) + x'(kn_{1} - m - 1)$ lie on a circle of radius $2\mbox{cos}(\frac{\pi (2m+1)}{2N_{2}})$ and let that circle be denoted by $C^{m}$. Then, $\mathcal{S}_{sum}$ takes the structure of $\frac{N_{2}}{2}$ concentric \textit{asymmetric} PSK signal sets. It is to be noted that $\mathcal{S}_{sum}$ takes the structure of $\frac{N_{2}}{2}$ concentric \textit{symmetric} PSK signal sets when $N_{1} = N_{2}$ \cite{HaR2}. Hence, the results of this paper is a non-trivial generalisation of the one in \cite{HaR2}. As an example, see Fig. \ref{sum_alphabet_pi_M} which shows the $\mathcal{S}_{sum}$ when $\mathcal{S}_{1}$ is a QPSK signal set and $\mathcal{S}_{2}$ is a $8$-PSK signal set. The set containing the radii of the $\frac{N_{2}}{2}$ circles is given by
\begin{equation*}
\mathcal{R} = \left\lbrace 2\mbox{cos}(\frac{\pi (2m + 1)}{2N_{2}}) ~|~ \forall ~ 0 \leq m \leq \frac{N_{2}}{2} - 1 \right\rbrace. 
\end{equation*}
\noindent Henceforth, throughout the section, $r(C^{m})$ denotes the radius of the circle $C^{m}$. Since the radius of each circle is a cosine function, the elements of $\mathcal{R}$ satisfies the relation, $$r(C^{M/2 - 1}) \leq r(C^{M/2 - 2}) \leq ~ \cdots ~ \leq r(C^{0}).$$ For the elements of $\mathcal{R}$, the following two propositions can be proved.
\begin{proposition}
\label{prop3}
The sequence $\left\lbrace r(C^{q}) - r(C^{q + 1}) \right\rbrace$ from $q = 0$ to $\frac{N_{2}}{2} - 2$ is an increasing sequence.
\end{proposition}
\begin{proof}
Using standard trigonometric identities, the term $r(C^{q}) - r(C^{q + 1})$ is given by $4\mbox{sin}(\frac{\pi}{N_{2}})\mbox{sin}(\frac{\pi (q+1)}{2N_{2}})$. Since $\frac{\pi (\frac{N_{2}}{2}-1)}{2N_{2}} \leq \frac{\pi}{2}$, the sequence $\{\mbox{sin}(\frac{\pi (q+1)}{2N_{2}})\}$ is an increasing sequence as $q$ increases from $0$ to $\frac{N_{2}}{2} - 2$.
\end{proof}
\begin{proposition}
\label{prop5}
Using the phase information of each point in $\mathcal{S}_{sum}$, the following observations can be made:
\begin{enumerate}

\item For a fixed $m$, the angular separation between the two points $x(n_{1}) + x'(kn_{1} + m)$ and $x(n'_{1}) + x'(kn'_{1} + m)$ on $C^{m}$ is $\frac{2 \pi (n_{1}-n'_{1})}{N_{1}}$ for all $m = 0 \mbox{ to } \frac{N_{2}}{2} - 1$. Similarly, for a fixed $m$, the angular separation between the two points $x(n_{1}) + x'(kn_{1} - m - 1)$ and $x(n'_{1}) + x'(kn'_{1} - m - 1)$ on $C^{m}$ is $\frac{2 \pi (n_{1}-n'_{1})}{N_{1}}$ for all $m = 0 \mbox{ to } \frac{N_{2}}{2} - 1$.
\item For a fixed $m$, the angular separation between the point, $x(n_{1}) + x'(kn_{1} + m)$ on $C^{m}$ and the point $x(n'_{1}) + x'(kn'_{1} - m - 1)$ on $C^{m}$ is $\frac{2 \pi (n_{1}-n'_{1})}{N_{1}} + \frac{\pi (2m + 1)}{N_{2}}$ for all $m = 0 \mbox{ to } \frac{N_{2}}{2} - 1$.
\item For a fixed $m$, the angular separation between the point $x(n_{1}) + x'(kn_{1} + m)$ on $C^{m}$ and the point $x(n'_{1}) + x'(kn'_{1} - (m-1) - 1)$ on $C^{m-1}$ is $\frac{2 \pi (n_{1}-n'_{1})}{N_{1}} + \frac{\pi (2m)}{N_{2}}$ for all $m = 1 \mbox{ to } \frac{N_{2}}{2} - 1$.
\item For a fixed $m$, the angular separation between the point $x(n_{1}) + x'(kn_{1} - m - 1)$ on $C^{m}$ and the point $x(n'_{1}) + x'(kn'_{1} + (m-1))$ on $C^{m-1}$ is $\frac{2 \pi (n_{1}-n'_{1})}{N_{1}} - \frac{\pi (2m)}{N_{2}}$ for all $m = 1 \mbox{ to } \frac{N_{2}}{2} - 1$.
\end{enumerate}
\end{proposition}

\indent In the next subsection, first, we partition both $\mathcal{S}_{1}$ and $\mathcal{S}_{2}$ into two groups (due to one bit redundancy added by the two encoders) using Ungerboeck rules and then, exploiting the structure of $\mathcal{S}_{sum}$, we compute the minimum Euclidean distance, $d_{min}$ of each one of the sets in $\mathcal{A}$.
\subsection{Structure of each one of the sets in $\mathcal{A}$ induced by the Ungerboeck partitioning on $\mathcal{S}_{1}$ and $\mathcal{S}_{2}$}
\label{sec3_subsec2}
For each $i = 1, 2$, let $\mathcal{S}_{i}$ be partitioned into two sets of equal size using Ungerboeck rules which results in two sets denoted by $\mathcal{S}^{e}_{i}$ and $\mathcal{S}^{o}_{i}$ such that $d_{min}$ of $\mathcal{S}^{e}_{i}$ and $\mathcal{S}^{o}_{i}$ is maximized. Since the number of sets resulting from the partition is two, the minimum angular separation, $\phi_{min}$ between the points in each set is $\frac{4\pi}{N_{1}}$. The two groups of $\mathcal{S}_{1}$ are of the form,
\begin{figure*}
\centering
\includegraphics[width=4in]{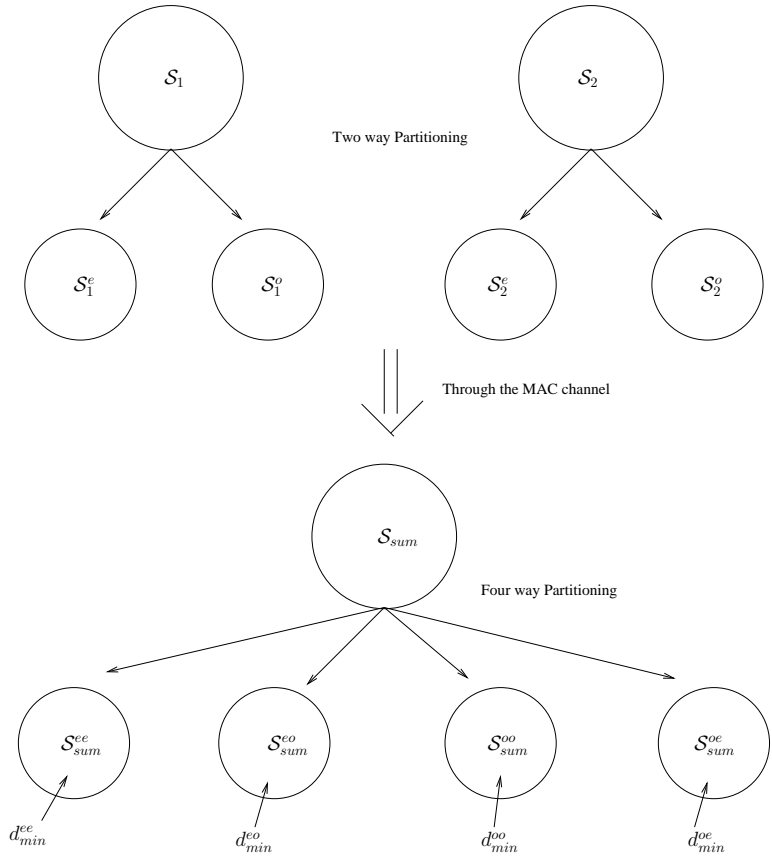}
\caption{Set partitioning of $\mathcal{S}_{sum}$ induced by the set partitioning of $\mathcal{S}_{1}$ and $\mathcal{S}_{2}$}
\label{partition}
\end{figure*}
\begin{equation*}
\mathcal{S}^{e}_1 = \left\lbrace x(n_{1}) ~|~ n_{1} = 2q ~ \mbox{ for }  0 \leq q \leq \frac{N_{1}}{2} - 1\right\rbrace \mbox{ and }
\end{equation*}
\begin{equation*}
\mathcal{S}^{o}_1 = \left\lbrace x(n_{1}) ~|~ n_{1} = 2q + 1 ~ \mbox{ for } 0 \leq q \leq \frac{N_{1}}{2} - 1\right\rbrace.
\end{equation*}
Similarly, the two groups of $\mathcal{S}_{2}$ are of the form,
\begin{equation*}
\mathcal{S}^{e}_2 = \left\lbrace x'(n_{2}) ~|~ n_{2} = 2l ~ \mbox{ for }  0 \leq l \leq \frac{N_{2}}{2} - 1\right\rbrace \mbox{ and }
\end{equation*}
\begin{equation*}
\mathcal{S}^{o}_2 = \left\lbrace x'(n_{2}) ~|~ n_{2} = 2l + 1 ~ \mbox{ for } 0 \leq l \leq \frac{N_{2}}{2} - 1\right\rbrace.
\end{equation*}
It is clear that the four sets $\mathcal{S}^{e}_1 + \mathcal{S}^{o}_2$, $\mathcal{S}^{e}_1 + \mathcal{S}^{e}_2$, $\mathcal{S}^{o}_1 + \mathcal{S}^{o}_2$ and $\mathcal{S}^{o}_1 + \mathcal{S}^{e}_2$ form a partition of $\mathcal{S}_{sum}$. Throughout this paper, the set $\mathcal{S}^{\alpha}_1 + \mathcal{S}^{\beta}_2$ and its minimum Euclidean distance are denoted by $S^{\alpha\beta}_{sum}$ and $d^{\alpha\beta}_{min}$ respectively $\forall~ \alpha, \beta = e$ and $o$. In the next subsection, only the structure of $S^{eo}_{sum}$ is studied and $d^{eo}_{min}$ value is calculated. The $d_{min}$ values of the rest of the sets in $\mathcal{A}$ can be calculated on the similar lines. The partition induced on $\mathcal{S}_{sum}$ due to the partition of $\mathcal{S}_{1}$ and $\mathcal{S}_{2}$ has been depicted in Fig. \ref{partition}.
\subsubsection{Calculation of $d_{min}$ of $\mathcal{S}^{e}_1 + \mathcal{S}^{o}_2$}
The elements of $\mathcal{S}^{e}_1 + \mathcal{S}^{o}_2$ (as given in \eqref{subset_1} at the top of the next page) are of the form $x(n_{1}) + x'(kn_{1} + m)$ and $x(n_{1}) + x'(kn_{1} - m - 1)$ where $n_{1}$ takes even value while $kn_{1} + m$ and $kn_{1} - m - 1$ take odd values. When $m$ is odd, note that $kn_{1} + m$ is odd and $kn_{1} - m - 1$ is even and hence $\mathcal{S}^{eo}_{sum}$ will have $\frac{N_{1}}{2}$ points of the form $x(n_{1}) + x'(kn_{1} + m)$ and no points of the form $x(n_{1}) + x'(kn_{1} - m - 1)$ on $C^{m}$. Similarly, when $m$ is even, $\mathcal{S}^{eo}_{sum}$ will have will have $\frac{N_{1}}{2}$ points of the form $x(n_{1}) + x'(kn_{1} - m - 1)$ and no points of the form $x(n_{1}) + x'(kn_{1} + m)$ on $C^{m}$. 

\indent Since $n_{1}$ takes only even values, using observation 1) of Proposition \ref{prop5}, $\phi_{min}$ between the points of $\mathcal{S}^{eo}_{sum}$ on any circle is $\frac{4\pi}{N_{1}}$. Hence the points of $\mathcal{S}^{eo}_{sum}$ are maximally separated on every circle. The following two propositions are important in finding the minimum Euclidean distance of the set $\mathcal{S}^{eo}_{sum}$.\\
\begin{figure*}
\begin{equation}
\label{subset_1}
\mathcal{S}^{e}_1 + \mathcal{S}^{o}_2 = \left\lbrace x(n_{1}) + x'(n_{2}) ~ | ~ n_{1} = 2q \mbox{ and } n_{2} = 2l + 1 ~\forall ~ q = 0 \mbox{ to } \frac{N_{1}}{2} -1 \mbox{ and } l = 0 \mbox{ to } \frac{N_{2}}{2} -1\right\rbrace.
\end{equation}
\hrule
\end{figure*}
\begin{proposition}
\label{prop6}
$r(C^{q-1})$ and $r(C^{q})$ satisfy the following inequality for all $q = 1$ to $\frac{N_{2}}{2} - 1$,
\begin{equation*}
d(r(C^{q-1}),  r(C^{q})e^{i\frac{2\pi}{M}}) \geq d(r(C^{q}), r(C^{q})e^{i\frac{2\pi}{N_{2}}}).
\end{equation*}
\end{proposition}
\begin{proof}
For $a, b \in \mathbb{C}$, let $\textit{l}(a, b)$ denote the line segment joining $a$ and $b$. Note that the three complex points $0, r(C^{q-1})$ and $r(C^{q-1})e^{i\frac{2\pi}{N_{2}}}$ form the three vertices of an isosceles triangle in $\mathbb{R}^{2}$. Since $r(C^{q}) \leq r(C^{q-1}),$ we have $d(0, r(C^{q})e^{i\frac{2\pi}{N_{2}}}) \leq  d(0, r(C^{q-1})e^{i\frac{2\pi}{N_{2}}})$. Therefore, the four points $r(C^{q}), r(C^{q})e^{i\frac{2\pi}{N_{2}}}, r(C^{q-1})$ and  $r(C^{q-1})e^{i\frac{2\pi}{N_{2}}}$ form the vertices of an isosceles trapezoid $\Upsilon$ such that $\textit{l}(r(C^{q}), r(C^{q})e^{i\frac{2\pi}{N_{2}}})$ is parallel to $\textit{l}(r(C^{q-1}), r(C^{q-1})e^{i\frac{2\pi}{N_{2}}})$. Also, note that $d(r(C^{q-1}),  r(C^{q})e^{i\frac{2\pi}{N_{2}}})$ is the length of the diagonal of the trapezoid $\Upsilon$. Since the angle between the line segments $\textit{l}(r(C^{q}), r(C^{q})e^{i\frac{2\pi}{N_{2}}})$ and $\textit{l}(r(C^{q}), r(C^{q-1}))$ is obtuse, $d(r(C^{q-1}),  r(C^{q})e^{i\frac{2\pi}{N_{2}}}) \geq d(r(C^{q}), r(C^{q})e^{i\frac{2\pi}{N_{2}}})$.
\end{proof}
\begin{proposition}
\label{prop7}
For $N_{1} \geq 8$, $r(C^{2k-1})$ satisfy the following inequality,
\begin{equation*}
d(r(C^{2k-1}), r(C^{2k-1})e^{i\frac{2\pi}{N_{2}}}) \geq 4\mbox{sin}(\frac{\pi}{N_{2}})\mbox{sin}(\frac{2\pi}{N_{1}}).
\end{equation*}
\end{proposition}
\begin{proof} Here, we prove the inequality $2r(C^{\frac{N_{2}}{2}-1})\mbox{sin}(\frac{2\pi}{N_{1}}) \leq d(r(C^{2k-1}) r(C^{2k-1})e^{i\frac{2\pi}{N_{2}}})$. The two terms across the inequality can be written as a ratio as shown below, 
\begin{equation} 
\label{exp2}
\frac{2r(C^{\frac{N_{2}}{2}-1})\mbox{sin}(\frac{2\pi}{N_{1}})}{d(r(C^{2k-1}), r(C^{2k-1})e^{i\frac{2\pi}{N_{2}}})} = \frac{\mbox{sin}(\frac{\pi}{2N_{2}})\mbox{sin}(\frac{4k\pi}{2N_{2}})}{\mbox{cos}(\frac{\pi (4k-1)}{2N_{2}})\mbox{sin}(\frac{\pi}{N_{2}})}.
\end{equation}
Since $\mbox{sin}(\frac{\pi}{2N_{2}}) < \mbox{sin}(\frac{\pi}{N_{2}})$, we have to prove that $\mbox{sin}(\frac{4k\pi}{2N_{2}}) < \mbox{cos}(\frac{\pi (4k-1)}{2N_{2}})$. Note that 
\begin{equation*}
\frac{\mbox{sin}(\frac{4k\pi}{2N_{2}})}{\mbox{cos}(\frac{\pi (4k-1)}{2N_{2}})} \leq \frac{\mbox{sin}(\frac{4k\pi}{2N_{2}})}{\mbox{cos}(\frac{\pi 4k}{2N_{2}})} = \mbox{tan}(\frac{\pi 4k}{2N_{2}}).
\end{equation*}
Note that $\mbox{tan}(\frac{\pi 4k}{2N_{2}}) \leq 1$ whenever $\frac{4k}{2N_{2}} \leq \frac{1}{4}$. The inequality $\frac{4k}{2N_{2}} \leq \frac{1}{4}$ holds when $N_{1} \geq 8$. This completes the proof.
\end{proof}

\indent Using the above two propositions, the $d_{min}$ value of the set $\mathcal{S}^{eo}_{sum}$ is presented in the following lemma.\\
\begin{lemma}
\label{lemma_1}
The minimum Euclidean distance of the set $\mathcal{S}^{eo}_{sum}$ is given by
{\small
\begin{equation}
\label{min_dist_set_1}
d^{eo}_{min} = 4\mbox{sin}(\frac{\pi}{2N_{2}})\mbox{sin}(\frac{2\pi}{N_{1}}).
\end{equation}
}
\end{lemma}
\begin{proof}
Since the points of $\mathcal{S}^{eo}_{sum}$ are maximally separated on every circle (with minimum angular separation of $\frac{4\pi}{N_{1}}$) and $C^{\frac{N_{2}}{2}-1}$ is the innermost circle, $d(r(C^{\frac{N_{2}}{2}-1}), r(C^{\frac{N_{2}}{2}-1})e^{i\frac{4\pi}{N_{1}}}) = 4\mbox{sin}(\frac{\pi}{2N_{2}})\mbox{sin}(\frac{2\pi}{N_{1}}) = d_{1}$ is a contender for $d^{eo}_{min}$. For this to be true, all other intra-distances in the set must be larger than or equal to $d_{1}$. In particular, we have to show that the distances between the points on any two consecutive circles must be larger than $d_{1}$.
In that direction, the first observation is the equality, $r(C^{2k-1}) - r(C^{2k}) = d_{1}$. From Proposition \ref{prop3}, $r(C^{q}) - r(C^{q+1}) \geq d_{1}$ for all $q \geq 2k-1$. Hence, the points on $C^{q}$ and $C^{q-1}$ (irrespective of their angular separation) are separated by a distance larger than $d_{1}$ for all $q = 2k$ to $\frac{N_{2}}{2} - 1$.\\
\indent Secondly, we must prove that a point on $C^{q}$ and a point on $C^{q-1}$ are separated by a distance larger than $d_{1}$ for all $q =1$ to $2k - 1$. In that direction, it can be shown that $\phi_{min}$ between a point on $C^{q}$ and a point on $C^{q-1}$ is one of the values from the set, $ \Omega = \{\pm \frac{2\pi}{N_{2}}, \pm \frac{4\pi}{N_{2}} \cdots \pm \frac{2(k-1)\pi}{N_{2}}\}$ depending on the value of $q$ for all $q =1$ to $2k - 1$. Using Proposition \ref{prop6}, we have 
\begin{equation*}
d(r(C^{2k-2}),  r(C^{2k-1})e^{i\frac{2\pi}{N_{2}}}) \geq d(r(C^{2k-1}),  r(C^{2k-1})e^{i\frac{2\pi}{N_{2}}}).
\end{equation*}
Furthermore, using the inequality in Proposition \ref{prop7}, we have
\begin{equation}
\label{inequal_1}
d(r(C^{2k-2}),  r(C^{2k-1})e^{i\frac{2\pi}{N_{2}}}) \geq d_{1}.  
\end{equation}
Note that the above inequality holds only when $N_{1} \geq 8$. However, when $N_{1} = 4$, we have $2k - 1 = \frac{N_{2}}{2} - 1$. Therefore, $r(C^{2k-1})= r(C^{\frac{N_{2}}{2}-1})$ will be the radius of the innermost circle wherein the minimum angular separation is $\frac{4 \pi}{N_{1}}$. Hence Proposition \ref{prop7} is not applicable when $N_{1} = 4$. Since $r(C^{q-1}) > r(C^{q})$, the inequality in \eqref{inequal_1} can be extended to
\begin{equation*}
d(r(C^{q-1}),  r(C^{q})e^{i\frac{2\pi}{N_{2}}}) \geq d_{1} \mbox{ for all } q =1 \mbox{ to } 2k - 1.
\end{equation*}
Hence, $d(r(C^{q-1}),  r(C^{q})e^{i\theta'}) \geq d_{1} \mbox{ for all } \theta' \in \Omega$. With this, we have proved that a point on $C^{q}$ and a point on $C^{q-1}$ are separated by a distance larger than $d_{1}$ for all $q =1$ to $2k - 1$. This completes the proof.
\end{proof}

\indent The $d_{min}$ values of the rest of the sets in $\mathcal{A}$ can be calculated on the similar lines. The following lemma provides the $d_{min}$ values of $\mathcal{S}^{oo}_{sum}$, $\mathcal{S}^{ee}_{sum}$ and $\mathcal{S}^{oe}_{sum}$.\\

\begin{lemma}
\label{lemma_2}
The minimum Euclidean distances of the sets $\mathcal{S}^{oo}_{sum}$, $\mathcal{S}^{ee}_{sum}$ and $\mathcal{S}^{oe}_{sum}$ are given by
{\small
\begin{equation}
\label{min_dist_set_2}
d^{ee}_{min} = d^{oe}_{min} = d^{oo}_{min}= 4\mbox{sin}(\frac{\pi}{2N_{2}})\mbox{sin}(\frac{2\pi}{N_{1}}).
\end{equation}
}
\end{lemma}
\begin{proof}
The proof is on the similar lines of the proof for Lemma \ref{lemma_1}.\\
\end{proof}
\subsection{Optimality of Ungerboeck partitioning}
\label{sec3_subsec3}
In the preceding subsection, $d_{min}$ values of each one of the sets of $\mathcal{A}$ induced by Ungerboeck partition on $\mathcal{S}_{1}$ and $\mathcal{S}_{2}$ have been computed in \eqref{min_dist_set_1} and \eqref{min_dist_set_2} (we refer $d^{U}_{min}$ = $d^{ee}_{min}$ = $d^{eo}_{min}$ = $d^{oo}_{min}$ = $d^{oe}_{min}$). In this subsection, using these values, it is shown that a non-Ungerboeck partition on either $\mathcal{S}_{1}$ or $\mathcal{S}_{2}$ results in a set $\mathcal{A}$ such that the $d_{min}$ of at least one of the sets in $\mathcal{A}$ is lesser than $d^{U}_{min}$.
\begin{theorem}
\label{ungerboeck_theorem}
For $\theta = \frac{\pi}{N_{2}}$, Ungerboeck partitioning on $\mathcal{S}_{1}$ and $\mathcal{S}_{2}$ into two sets is optimal in maximizing the minimum of the $d_{min}$ values of the sets in $\mathcal{A}$.
\end{theorem}
\begin{proof}
Let $\mathcal{S}^{1}_{i}$ and $\mathcal{S}^{2}_{i}$ be the two sets (of equal cardinality) resulting from a partition of $\mathcal{S}_{i}$ for $i = 1, 2$. If either $\mathcal{S}_{1}$ or $\mathcal{S}_{2}$ is not Ungerboeck partitioned, then it is to be shown that, $d_{min}$ of at least one of the sets in the set $\mathcal{A} = \left\lbrace \mathcal{S}^{i}_{1} + \mathcal{S}^{j}_{2} ~ |~ \forall ~ i, j = 1, 2 \right\rbrace$ is lesser than $d^{U}_{min}$. Here, we prove the above result when $\mathcal{S}_{2}$ is not Ungerboeck partitioned. On the similar lines, the above result can be proved when $\mathcal{S}_{1}$ is not Ungerboeck partitioned as well. We show that $\phi_{min}$ between any two points in one of sets in $\mathcal{A}$ is smaller than $\frac{4\pi}{N_{1}}$ on the circle $C^{\frac{N_{2}}{2} -1}$. It is assumed that there are exactly $\frac{N_{1}}{2}$ points on $C^{\frac{N_{2}}{2} - 1}$ in each set of $\mathcal{A}$. Otherwise, at least one set contains more than $\frac{N_{1}}{2}$ points on $C^{\frac{N_{2}}{2} - 1}$ and hence $\phi_{min}$ between a pair of points in that set will be lesser than $\frac{4\pi}{N_{1}}$. Therefore, the sub-optimality of the partition can be proved.  Hence, without loss of generality, we assume that $x'(a), x'(a+ 1) \in \mathcal{S}^{1}_{2}$ for some $a$ such that $0 \leq a \leq N_{2}-2$. Note that the elements of $\mathcal{S}_{2}$ are of the form $x'(kn_{1} + m)$ or $x'(kn_{1} - m - 1)$. Since $ m = \frac{N_{2}}{2} - 1$ is odd and $(a, a + 1)$ is an even and odd pair, $x'(a)$ can be of the form $x'(kn_{1} + m)$ and $x'(a+ 1)$ can be of the form $x'(kn_{1} - m - 1)$ or vice-verse. Without loss of generality, we assume that $x'(a)$ is of the form $x'(kn_{1} + \frac{N_{2}}{2} -1)$ and $x'(a+ 1)$ is of the form $x'(kn'_{1} - \frac{N_{2}}{2})$ for some $n_{1}$, $n'_{1}$. Note that, the points $x(n_{1}) + x'(kn_{1} + \frac{N_{2}}{2} -1)$ and $x(n'_{1}) + x'(kn'_{1} - \frac{N_{2}}{2})$ belong to one of the sets in $\mathcal{A}$ and have an angular separation of $\frac{2\pi(n_{1} - n'_{1})}{N_{1}} + \frac{(N_{2} - 1)\pi}{N_{2}}$. This implies that there exists a pair of points on $C^{\frac{N_{2}}{2} - 1}$ such that $\phi_{min}$ between them is lesser than $\frac{4\pi}{N_{1}}$. This completes the proof.
\end{proof}

\indent From the above theorem, it is clear that Ungerboeck labelling on the trellis of each user can be used systematically to construct trellis code pairs to approach \textit{any rate pair} within the CC capacity region.
\subsection{On the choice of the cardinality of the signal sets}
\label{sec3_subsec4}
Using the results presented in the preceding subsection, we illustrate how to choose the cardinality of PSK signal sets to achieve any rate pair $(r_{1}, r_{2})$ (assuming $r_{2} \geq r_{1}$) within the sector O-A-B shown in Fig. \ref{CC_region_equal}. We do not consider achieving rate pairs outside the sector O-A-B since such points can be moved either horizontally or vertically (or both) into the sector O-A-B which in-turn either increases the rate for both users or increases the rate for one of the users by keeping the rate for the other as it is. For example, achieving the rate pair denoted by P' shown in Fig. \ref{CC_region_equal} is better than achieving the rate pair denoted by P since the latter has higher sum-rate by retaining the same rate for user-2 and increasing the rate for user-1.\\
\indent For a given equal power constraint, to approach a rate pair $(r_{1}, r_{2})$, we choose sufficiently large values of $N_{1}$ and $N_{2}$ such that
\begin{enumerate}
\item $N_{1}$ and $N_{2}$ satisfies the following approximation,
\begin{equation*}
\frac{\mbox{log}_{2}(N_{2})}{\mbox{log}_{2}(N_{1})} \simeq  \frac{r_{2}}{r_{1}} \mbox{ and }
\end{equation*} 
\item the CC capacity region with $N_{1}$-PSK and $N_{2}$-PSK signal sets encloses the point $(r_{1}, r_{2})$.
\end{enumerate}

\indent For the above choice of ($N_{1}, N_{2}$), if the number of input bits for user-$i$ is $\mbox{log}_{2}(N_{i}) - 1$, and Ungerboeck labelling is employed on the trellis (with larger number of states) of each user, then the rate pair $(r_{1}, r_{2})$ can be approached. To approach larger rates (close to the line segment A-B) on the same line segment, larger values of $N_{1}$ and $N_{2}$ have to be chosen satisfying according to points 1) and 2) mentioned above. Therefore, with sufficiently large values of $N_{1}$ and $N_{2}$ satisfying the conditions 1) and 2), any pair, $(r_{1}, r_{2})$ in the Gaussian capacity region can be approached.  

\section{Discussion}
\label{sec4}
In this paper, we have designed the optimal labelling rule on the trellises of the two users (with equal average power constraints on the two users) to approach any rate pair within the capacity region. Some possible directions for further work are as follows:  
\begin{itemize}
\item In this paper, we have assumed equal average power constraint for both the users. If unequal power constraints are considered, then it is interesting to see if rotation between the alphabets provides enlargement of the CC capacity region.
\item We have designed TCM schemes with unequal rates for the two users even though the two users have equal average power constraints. However, when rates proportional to the power constraint of each user has to be transmitted (with unequal average power constraints on the two users), TCM schemes have to be designed with pairs of signal sets which have unequal power constraints and unequal size.
\end{itemize}
\section*{Acknowledgement}
This work was partly supported by the DRDO-IISc Program on Advanced Research in Mathematical Engineering to B. S. Rajan.

\end{document}